\pdfoutput=1
\documentclass[conference]{IEEEtran}
\usepackage{cite}
\usepackage{marvosym}
\usepackage[utf8]{inputenc}
\usepackage[table]{xcolor}
\usepackage{amsmath}
\usepackage{multirow}
\usepackage{booktabs}
\usepackage{adjustbox}
\usepackage[english]{babel}
\usepackage{graphicx}
\usepackage{tcolorbox}
\usepackage{enumitem}
\usepackage{makecell}
\usepackage{threeparttable}
\usepackage{hyperref}
\usepackage{array}
\usepackage{balance}
\usepackage{etoolbox}
\makeatletter
\patchcmd{\@makecaption}{\scshape}{\upshape}{}{\PackageError{captionstyle}{Could not update table caption font}{}}
\makeatother

\newtcolorbox{takeaway}[1][]{
  colback=gray!8,
  colframe=gray!65!black,
  coltitle=white,
  fonttitle=\bfseries\small,
  title=Takeaway,
  boxrule=0.4pt,
  arc=2pt,
  left=5pt,right=5pt,top=3pt,bottom=3pt,
  before skip=4pt,after skip=4pt,
  #1
}

\author{
    \IEEEauthorblockN{
        Junda He\IEEEauthorrefmark{1},
        Xiangyu Shan\IEEEauthorrefmark{2},
        Jieke Shi\IEEEauthorrefmark{1},
        Yunbo Lyu\IEEEauthorrefmark{1}\textsuperscript{\,\Letter},
        Gou Tan\IEEEauthorrefmark{3}, \\
        Xiuwei Shang\IEEEauthorrefmark{1},
        Zhou Yang\IEEEauthorrefmark{4},
        Christoph Treude\IEEEauthorrefmark{1}, and
        David Lo\IEEEauthorrefmark{1}
    }

    \IEEEauthorblockA{\IEEEauthorrefmark{1}Singapore Management University \\
    jundahe.2022@phdcs.smu.edu.sg, \{jiekeshi, yunbolyu, xwshang, ctreude, davidlo\}@smu.edu.sg}

    \IEEEauthorblockA{\IEEEauthorrefmark{2}Yangzhou University \\
    231305205@stu.yzu.edu.cn}

    \IEEEauthorblockA{\IEEEauthorrefmark{3}School of Systems Science and Engineering, Sun Yat-sen University, Guangzhou, 510006, P.R. China \\
    tang29@mail2.sysu.edu.cn}

    \IEEEauthorblockA{\IEEEauthorrefmark{4}University of Alberta; Alberta Machine Intelligence Institute; Canada CIFAR AI Chair \\
    zy25@ualberta.ca}
}

\newcommand{\phead}[1]{%
  \par\addvspace{0.4\baselineskip}%
  \noindent\textbf{#1}\hskip 0.75em\ignorespaces}

\begin{document}

\title{When Agreement Hides Disagreement:\\ Are We Evaluating LLM Judges Fairly?}

\maketitle
\thispagestyle{plain}
\pagestyle{plain}
\raggedbottom
\setlength{\textfloatsep}{18pt plus 2pt minus 2pt}
\setlength{\dbltextfloatsep}{18pt plus 2pt minus 2pt}

\begin{abstract}
Software engineering (SE) studies often evaluate large language model (LLM) judges by measuring agreement with a single pooled human label. When human raters disagree, interpreting that score requires knowing how closely a human would agree with the rest of the panel. Prior work in NLP observed that judges can appear to outperform humans on the items humans dispute most. We examine how much of this apparent advantage persists under a comparable human baseline. We introduce \textbf{Consensus-Sensitive Judge Evaluation (CSJE)}, which groups items by human agreement and compares the judge with a held-out human within each group. Across three benchmarks covering code summarization, commit message generation, and code readability, the best judges reported by the original studies score below the matched human baseline overall. On contested commit messages, matching the baseline reduces GPT-4's apparent advantage in judging how well a message explains a change. On contested code summaries, a human-only control reproduces GPT-4-turbo's apparent advantage in judging content adequacy. These results show how pooling human ratings and selecting items by disagreement can affect conclusions about judge performance. Three frontier models (Claude Opus~4.8, Gemini~3.1~Pro, and GPT-5.5) and two multi-agent methods (CodeVisionary and SE-Jury) yield no consistent gain over the released judges. For comparisons that require a single score, we introduce Human-Ceiling Agreement (HCA), which summarizes proximity to individual ratings relative to random and human references. We recommend reporting this summary alongside performance and matched human agreement in each consensus stratum.
\end{abstract}


\section{Introduction}\label{sec:intro}

Recently, Large Language Models (LLMs) have demonstrated great capability in solving complex software engineering (SE) tasks~\cite{hou2024large,xia2023automated,zhang2024autocoderover}. Inspired by these advancements, the SE community has shown growing interest in the LLM-as-a-Judge paradigm~\cite{he2026llm,crupi2025effectiveness,zhou2025se,zeng2025evaluating,wang2025can}, which shifts the role of LLMs from generating software artifacts to evaluating artifacts. As a cost-effective and highly scalable alternative to manual evaluation, LLM judges have been widely applied to assess the quality of source code~\cite{ouedraogo2025coreeval}, code summaries~\cite{crupi2025effectiveness,zhou2025se} and commit messages~\cite{zeng2025evaluating}.

Ultimately, we want LLM judges to be able to evaluate software artifacts in place of human evaluators. To serve this role, their judgments should reflect human judgment. A common practice in SE studies~\cite{crupi2025effectiveness,zhou2025se,zeng2025evaluating,wang2025can} is to combine human ratings for each software artifact into a single value, which we call the \textit{pooled label}. Researchers typically take the median or mean of the ratings and treat the resulting label as the ground-truth reference. They then measure agreement between the judge's ratings and the pooled labels using metrics such as Krippendorff's~$\alpha$~\cite{krippendorff2018content} or Spearman's~$\rho$~\cite{spearman1904proof}. However, we question whether this practice alone tells us how well an LLM judge reflects human judgment. This evaluation practice has two limitations, one at the \textit{item level} and the other at the \textit{benchmark level}.

The first limitation concerns how human disagreement is handled at the \textit{item level}. Many SE criteria are subjective, e.g., the adequacy or conciseness of a code summary. Human ratings can naturally diverge, due to individual subjectivity, bias, or uncertainty~\cite{tversky1974judgment}; therefore some disagreement is unavoidable whenever humans evaluate~\cite{kahneman2021noise}.
We argue that for subjective SE criteria, a single ground-truth rating may not exist at all. However, prior SE judge studies usually treat the pooled label as ground truth, overlooking the disagreement in the human judgments. We call this assumption the \textit{ground-truth illusion}: treating an aggregate of potentially conflicting judgments as a uniquely correct reference.

The second limitation concerns how judge performance is summarized at the \textit{benchmark level}. Prior studies report a single agreement coefficient over the entire benchmark, regardless of how contested its individual items are. Such a number blends items on which annotators agree almost perfectly with items on which they barely agree, so a healthy-looking overall score can hide wide variation underneath.

Elangovan et al.~\cite{elangovan2025beyond} address the benchmark-level problem by stratifying items by annotator agreement and evaluating the judge within each stratum. 
This reveals how judge agreement varies across different levels of human consensus.
The item-level limitation remains, however, because the pooled label still serves as ground truth within each stratum. On contested items, several ratings may be defensible, so a judge can give a reasonable rating that differs from the pooled label. To interpret the judge's agreement with that label, we need to know how closely an individual rater agrees with the pooled ratings of the remaining annotators on the same items.

To our knowledge, no prior SE study has examined how judge performance compares with this human reference across levels of annotator agreement. We propose \textit{Consensus-Sensitive Judge Evaluation} (CSJE) to make this comparison. First, CSJE groups items by annotator agreement. Second, within each stratum, we measure the judge's agreement with the pooled label, as in prior evaluations. To interpret this score, we compare it with a held-out human rater's agreement, and we also treat the judge as an additional rater. We further use controls to check how selecting items by human disagreement affects these comparisons. Third, when a single score is needed, Human-Ceiling Agreement (HCA) summarizes how closely the judge agrees with individual raters. Random scoring defines $0$ on this scale, and agreement among human raters defines the human reference of $1$. Items receive more weight when raters agree more closely with one another than a random scorer would.

We apply CSJE to three SE tasks: code summarization~\cite{crupi2025effectiveness}, commit message generation~\cite{zeng2025evaluating}, and code readability assessment~\cite{ouedraogo2025coreeval}. We first measure how common human disagreement is on each benchmark. We then examine how this disagreement affects the evaluation of LLM judges. We extend this comparison to three frontier LLMs (Claude Opus~4.8, Gemini~3.1~Pro, and GPT-5.5) and two state-of-the-art multi-agent judging methods (CodeVisionary~\cite{wang2025codevisionary} and SE-Jury~\cite{zhou2025se}).

Across the three benchmarks, human disagreement is widespread, and all best-reported judges from the original studies fall below the matched human baseline overall. Where judges appear to outperform humans on contested items, our matched baseline and human-only controls show that pooling ratings or selecting items by disagreement can produce the apparent advantage, so it does not establish superior judging. Moreover, neither the frontier LLMs nor the multi-agent methods consistently improve upon the originally reported judges.

\noindent Our major contributions are as follows:
\begin{itemize}[leftmargin=*]
    \item Across three SE benchmarks, we measure how often human raters disagree and evaluate how judge agreement varies with human consensus across the original best-reported judges, three frontier LLMs, and two multi-agent methods.
    \item We propose the Consensus-Sensitive Judge Evaluation (CSJE) protocol. CSJE compares judges with a held-out human within each consensus stratum and uses controls to assess how selecting items by disagreement affects that comparison.
    \item We introduce Human-Ceiling Agreement (HCA), a metric that summarizes judge quality while accounting for disagreement among human annotators.
\end{itemize}
\section{Study Design}\label{sec:method}

This section details the benchmarks and our Consensus-Sensitive Judge Evaluation (CSJE) protocol: how we stratify items by human consensus and how we measure judge agreement.

\subsection{Benchmarks and Evaluated LLM Judges}\label{subsec:benchmarks}

\noindent We base our study on three prior SE works that evaluate LLM judges by their agreement with human ratings~\cite{crupi2025effectiveness,zeng2025evaluating,ouedraogo2025coreeval}, and we re-analyze each using its released multi-annotator benchmark and LLM-judge scores (summarized in Table~\ref{tab:benchmarks}). Together they cover three distinct tasks: code summarization~\cite{crupi2025effectiveness}, commit message generation~\cite{zeng2025evaluating}, and code readability assessment~\cite{ouedraogo2025coreeval}. For each benchmark we take the best judge's released scores from its official replication package as our primary analysis target.

\begin{table}[t]
  \centering
  \caption{Benchmark characteristics and evaluated LLM judges.}
  \label{tab:benchmarks}
  \footnotesize
  \setlength{\tabcolsep}{3pt}
  \adjustbox{max width=\columnwidth}{
  \begin{tabular}{l c c c l l}
    \toprule
    \textbf{Benchmark} & \textbf{Items} & \textbf{Raters} & \textbf{Scale} & \textbf{Judge (Prompt)} & \textbf{Criteria} \\
    \midrule
    \makecell[l]{Code Summarization\\\cite{crupi2025effectiveness}} & 1{,}163 & 3 & 1--5 & \makecell[l]{GPT-4-turbo\\(Zero-shot)} & \makecell[l]{Content Adequacy\\Conciseness\\Fluency} \\
    \midrule
    \makecell[l]{Commit Msg. Gen.\\\cite{zeng2025evaluating}} & 400 & 6 & 0--4 & \makecell[l]{GPT-4\\(Few-shot+CoT)} & \makecell[l]{What\\Why} \\
    \midrule
    \makecell[l]{Code Readability\\\cite{ouedraogo2025coreeval}} & 124 & 9 & 1--5 & \makecell[l]{Mistral Large\\(Zero-shot)} & Readability \\
    \bottomrule
  \end{tabular}
  }
\end{table}

\noindent\textbf{Code Summarization.} We use the dataset from Crupi
et al.~\cite{crupi2025effectiveness} (derived from CoderEval~\cite{yu2024codereval}). It contains $1{,}163$ code--summary pairs across Java and Python.
Three of nine annotators rated each summary
on a 1--5 scale across three dimensions: Content Adequacy, Conciseness, and Fluency.
The released data list the three ratings of a summary without annotator identities, so a rater index on this benchmark denotes a rating column, not a fixed person.
The source study evaluates several LLM judges (e.g., GPT-3.5-turbo, GPT-4-turbo, and CodeLlama-7B/13B/34B) and reports GPT-4-turbo (zero-shot) as the best-aligned, which we adopt.

\noindent\textbf{Commit Message Generation.} We adopt the benchmark from Zeng et al.~\cite{zeng2025evaluating}, with 400 human-written and 400 LLM-generated commit messages. Six volunteers independently rated each message on a 0--4 Likert scale (0 = poor, 4 = excellent) along two dimensions: \textit{What} (how well it captures the code changes) and \textit{Why} (how well it explains the reason behind them). Following Zeng et al., we use the 400 human-written messages, which we refer to as the \textit{developer pool}, and adopt GPT-4 (Few-shot+CoT), the configuration they report as the best among four LLM judges and prompting strategies.

\noindent\textbf{Code Readability Assessment.} We use the Scalabrino panel~\cite{scalabrino2018comprehensive} from the CoReEval readability benchmark~\cite{ouedraogo2025coreeval}, with 124 method-level Java snippets from real-world projects, each rated for readability by nine computer science students on a 5-point Likert scale. The source study evaluates ten LLM judges (four API-based and six open-source) and identifies Mistral~Large (zero-shot) as the best-aligned API model.

\subsection{Measuring Human Consensus}\label{subsec:human-uncertainty}

\noindent Instead of a single agreement coefficient for the whole dataset, we measure consensus on each item separately and assign it to one of three strata, \textit{High}, \textit{Partial}, or \textit{Low}, using two metrics:
\begin{itemize}[leftmargin=*]
\item \textbf{Majority Agreement (MA)}~\cite{elangovan2025beyond}: the fraction of annotators whose score matches the panel's median or mode for that item. Higher MA indicates stronger agreement.
\item \textbf{Range}: The absolute difference between the maximum and minimum scores assigned to an item by the annotator panel. A Range of 0 indicates perfect unanimity.
\end{itemize}

\noindent For all three tasks, we use the median of the human ratings for each item as its pooled human label, following prior work~\cite{crupi2025effectiveness}.

Because the three benchmarks differ in panel size (three to nine raters) and rating scale, we set the stratum thresholds separately for each task:

\begin{itemize}[leftmargin=*]
\item \textit{Code Summarization (Three raters)}. We use MA, which gives three strata: \textit{High} ($\text{MA}=1.0$; all three match), \textit{Partial} ($\text{MA}=2/3$; two match), and \textit{Low} ($\text{MA}<2/3$; all differ).
\item \textit{Commit Message Generation (Six raters)}. With an even number of raters, the median often lands on a half-integer, making MA ill-defined. We thus use Range as the primary metric: \textit{High} ($\text{Range}=0$), \textit{Partial} ($\text{Range}=1$), and \textit{Low} ($\text{Range}\ge 2$).
\item \textit{Code Readability Assessment (Nine raters)}. A nine-rater panel almost never reaches perfect unanimity ($\text{MA}=1.0$ or $\text{Range}=0$). We therefore use a relaxed MA threshold based on the median: \textit{High} is a strict majority ($\text{MA}\ge 5/9$), \textit{Partial} is $3/9\le\text{MA}<5/9$, and \textit{Low} is $\text{MA}<3/9$.
\end{itemize}

We report the resulting item proportions and human--human agreement in RQ1 (Table~\ref{tab:human_consensus}).

\subsection{Agreement Metrics}\label{subsec:metrics}

\noindent Following prior work, we first evaluate the LLM judge against the pooled human label, assuming a single ground-truth rating exists for each item. Within each consensus stratum, we compare the judge's agreement with the pooled labels against agreement among human raters on the same items.

For $N$ items rated by $K$ human annotators, let $h_{ij}$ denote rater $j$'s score on item $i$ ($j=1,\dots,K$), $\tilde{h}_i = \operatorname{median}_{j}(h_{ij})$ the pooled human label, and $l_i$ the judge's score. Letting $\alpha(\cdot)$ denote ordinal Krippendorff's $\alpha$~\cite{krippendorff2018content}, we define human--human agreement and judge--pooled agreement as
\begin{equation}
\begin{aligned}
\alpha_{\text{H-H}} \;&=\; \alpha\!\left(\{(h_{i1}, \dots, h_{iK})\}_{i=1}^{N}\right), \\
\alpha_{\text{H-L}} \;&=\; \alpha\!\left(\{(\tilde{h}_i, l_i)\}_{i=1}^{N}\right).
\end{aligned}
\label{eq:alpha-defs}
\end{equation}
Prior studies~\cite{crupi2025effectiveness,zeng2025evaluating,elangovan2025beyond} measure the gap between LLM judges and human raters by comparing these two scores, which we formalize as the \textit{raw gap}:
\begin{equation}
\Delta_\alpha^{\text{raw}} \;=\; \alpha_{\text{H-H}} \;-\; \alpha_{\text{H-L}}.
\label{eq:agreement-gap}
\end{equation}

However, this raw gap is confounded by two factors. First, the pooled human label denoises the target, meaning the judge is evaluated against a smoothed, idealized baseline while humans are compared against noisy individual ratings. Second, selecting items on which human raters disagree tends to lower measured human agreement. It is therefore hard to tell how much of the raw gap on these items reflects the judge's behavior.

To make the comparison fairer, we introduce two complementary measures: a \textit{matched gap} that also scores humans against a pooled target, and a \textit{pairwise gap} that treats the judge as an additional rater without pooling.

\phead{Matched Gap.} To construct a comparable human baseline, the \textit{held-out-human reference}~\cite{rajpurkar2016squad} holds every rater out in turn. Note that we do not randomly select just one rater as the reference. Let $\tilde{h}_{i,-j} = \operatorname{median}_{k \neq j}(h_{ik})$ denote the pooled label of the remaining $K{-}1$ raters on item $i$. Pairing each rater's scores with $\tilde{h}_{i,-j}$ across all $N$ items and $K$ raters yields $KN$ rating pairs, over which we compute a single ordinal $\alpha$:
\begin{equation}
\alpha_{\text{ref}} \;=\; \alpha\!\left(\{(\tilde{h}_{i,-j}, h_{ij})\}_{1\le i\le N,\, 1\le j\le K}\right),
\label{eq:alpha-ref}
\end{equation}
which gives the \textit{matched gap}:
\begin{equation}
\Delta_\alpha^{\text{match}} \;=\; \alpha_{\text{ref}} \;-\; \alpha_{\text{H-L}}.
\label{eq:matched-gap}
\end{equation}
Both agents are now scored against a comparably denoised target. The targets still differ in panel size: the judge is scored against the $K$-rater median, whereas each held-out human faces the $(K{-}1)$-rater median. This difference may affect target stability and should be considered when interpreting the matched gap. Section~\ref{sec:threats} reports a check that removes it. The raw gap decomposes exactly as $\Delta_\alpha^{\text{raw}} = \Delta_\alpha^{\text{match}} + (\alpha_{\text{H-H}} - \alpha_{\text{ref}})$, where the second term is the part due to aggregation.

\phead{Selection Controls.} Even after matching, a post-selection bias (or ``double dipping'') remains because the same human ratings $h_{ij}$ are used both to assign items to consensus strata and to compute the human baseline $\alpha_{\text{ref}}$, whereas the judge's scores $l_i$ play no role in item selection~\cite{kriegeskorte2009circular,taylor2015statistical}. Specifically, items enter the \textit{Low} stratum precisely because the observed human ratings diverge on them, whether from genuine ambiguity or rater noise. Conditioning on this disagreement mechanically depresses $\alpha_{\text{ref}}$ among those raters, while an independent scorer not involved in selection is not subject to that constraint and regresses toward the panel center~\cite{barnett2005regression}, which can produce an apparent judge advantage ($\Delta_\alpha^{\text{match}}<0$).
To isolate this selection effect, we use two controls: the \textit{split-panel control}, which decouples stratum selection from baseline estimation using disjoint rater groups, and the \textit{constant-quality null}, which places a held-out human rater in the judge's role outside item selection.

\begin{itemize}[leftmargin=*]
\item \textbf{Split-panel control.}
Our design requires $K\ge6$ raters because it uses two disjoint groups of three.
For commit message generation, the first group assigns items to strata using the Range rule.
Within each stratum, we compute $\alpha_{\text{ref}}$ by holding out each rater in the second group in turn and comparing their ratings with the item-wise median of the other two.
We score the LLM judge against the item-wise median of all three raters in the second group.
We report the mean matched gap over all 20 choices of the stratifying group.

\item \textbf{Constant-quality null.}
We hold out each of the $K$ raters in turn to act as the judge.
The remaining $K{-}1$ raters define the strata and provide both $\alpha_{\text{ref}}$ and the pooled labels used to score the held-out rater.
Within each stratum, we average the matched gap over the $K$ choices of judge.
A negative gap shows that selecting items by panel disagreement can produce an apparent advantage even for a human judge.
\end{itemize}

The matched reference requires at least four raters for effective denoising.
On the three-rater code-summarization panel, $\alpha_{\text{ref}}\approx\alpha_{\text{H-H}}$ in most cases, so we rely on the constant-quality null alone to assess selection effects.
In this control, an item is contested when the two remaining raters disagree.

\phead{Judge as an Additional Rater.}
Instead of pooling human ratings into a single label, we also treat the LLM judge as an additional rater alongside the $K$ human raters and measure pairwise agreement directly~\cite{williams1976comparing}. For every pair of distinct human raters $(j,k)$, we compute ordinal $\alpha$ between their scores across the $N$ items and average across all human--human pairs to obtain $\alpha_{\text{H-H}}^{\text{pair}}$. Similarly, for each human rater $j$, we compute ordinal $\alpha$ between their scores and the judge's scores, and average across all $K$ judge--human pairs to obtain $\alpha_{\text{J-H}}^{\text{pair}}$:
\begin{equation}
\begin{aligned}
\alpha_{\text{H-H}}^{\text{pair}} \;&=\; \frac{1}{\binom{K}{2}} \sum_{1\le j<k\le K} \alpha\!\left(\{(h_{ij}, h_{ik})\}_{i=1}^{N}\right), \\
\alpha_{\text{J-H}}^{\text{pair}} \;&=\; \frac{1}{K} \sum_{j=1}^{K} \alpha\!\left(\{(h_{ij}, l_i)\}_{i=1}^{N}\right),
\end{aligned}
\label{eq:pairwise-defs}
\end{equation}
and report their difference as the \textit{pairwise gap} over all items and within each consensus stratum:
\begin{equation}
\Delta^{\text{pair}} \;=\; \alpha_{\text{H-H}}^{\text{pair}} \;-\; \alpha_{\text{J-H}}^{\text{pair}}.
\label{eq:pairwise-gap}
\end{equation}

\phead{Significance and multiple comparisons.}
To test whether each $\Delta_\alpha$ is distinguishable from sampling noise, we resample items with replacement ($B{=}2000$)~\cite{10.1214/aos/1176344552} and recompute $\Delta_\alpha$ in each bootstrap sample. For each stratified comparison, we calculate the proportions of bootstrap gaps that are nonnegative and nonpositive. The two-sided sign-bootstrap $p$-value is twice the smaller proportion, capped at 1. Because each table contains multiple such comparisons, we apply Benjamini--Hochberg~\cite{benjamini1995controlling} at $q<0.05$ within each table to control the false discovery rate.

\section{Results}\label{sec:results}

\subsection*{RQ1: How much human uncertainty do current SE judge benchmarks hide?}\label{sec:rq1}

\noindent We examine how often human raters disagree on three benchmarks used in prior studies to evaluate LLM judges for software engineering tasks. A single agreement coefficient for the entire benchmark can hide whether disagreement affects a few items or many. For each rating criterion used in the original study, we group items by their level of human consensus (\textit{High}, \textit{Partial}, or \textit{Low}). We report the percentage of items and the human--human agreement within each group (Table~\ref{tab:human_consensus}).

\begin{table*}[t]
  \centering
  \caption{Human--human agreement (ordinal Krippendorff's $\alpha$) by consensus stratum; item shares are in parentheses. Code-summarization $\alpha$ values are averaged over Java and Python.}
  \label{tab:human_consensus}
  \setlength{\tabcolsep}{6pt}
  \adjustbox{max width=\textwidth}{
  \begin{tabular}{l l l c c c c}
    \toprule
    \textbf{Benchmark} & \textbf{Criterion} & \textbf{Language} & \textbf{Overall} & \textbf{High} & \textbf{Partial} & \textbf{Low} \\
    \midrule
    \multirow{6}{*}{Code Summarization} & \multirow{2}{*}{Content Adequacy} & Java & \multirow{2}{*}{0.75} & 1.00 (51\%) & 0.54 (45\%) & 0.01 (4\%) \\
     &  & Python &  & 1.00 (32\%) & 0.54 (56\%) & 0.01 (12\%) \\
    \addlinespace[3pt]
     & \multirow{2}{*}{Conciseness} & Java & \multirow{2}{*}{0.52} & 1.00 (64\%) & 0.13 (33\%) & -0.23 (3\%) \\
     &  & Python &  & 1.00 (43\%) & 0.13 (51\%) & -0.23 (6\%) \\
    \addlinespace[3pt]
     & \multirow{2}{*}{Fluency} & Java & \multirow{2}{*}{0.50} & 1.00 (48\%) & 0.27 (47\%) & -0.21 (5\%) \\
     &  & Python &  & 1.00 (30\%) & 0.27 (60\%) & -0.21 (10\%) \\
    \midrule
    \multirow{2}{*}{Commit Msg. Gen.} & What & & 0.57 & 1.00 (8.5\%) & 0.76 (46.5\%) & 0.24 (45.0\%) \\
    & Why & & 0.69 & 1.00 (21.2\%) & 0.83 (28.8\%) & 0.40 (50.0\%) \\
    \midrule
    Code Readability & Readability & & 0.23 & 0.35 (48\%) & 0.12 (45\%) & -0.02 (7\%) \\
    \bottomrule
  \end{tabular}
  }
\end{table*}

\phead{Code Summarization.} Overall, human--human agreement looks healthy (Krippendorff's $\alpha=0.50$--$0.75$ across the three criteria). Once items are split into consensus strata, agreement falls monotonically. The \textit{High} stratum, where all three annotators gave the same score (so $\alpha=1.00$), covers 30--64\% of items. The \textit{Partial} stratum drops to $\alpha=0.13$--$0.54$. In the \textit{Low} stratum, agreement collapses to near zero or below. This contested region is only a small tail (3--12\% of items).

\phead{Commit Message Generation.} The same monotonic collapse appears. The \textit{High} stratum, where all six raters gave the same score (so $\alpha=1.00$ by definition), covers only 8.5\% of items for \textit{What} and 21\% for \textit{Why}. The \textit{Partial} stratum stays high ($\alpha=0.76$--$0.83$). But the \textit{Low} stratum covers nearly half of all items (45--50\%) at only $\alpha=0.24$--$0.40$. Thus, the overall agreement score masks substantial disagreement on nearly half of the commit messages.

\phead{Code Readability Assessment.} This benchmark is rated by nine annotators, and it is the most internally divided benchmark: the overall human--human agreement is only $\alpha=0.23$. The \textit{High} stratum reaches only $\alpha=0.35$; the \textit{Partial} stratum (45\% of items) collapses to $\alpha=0.12$.

\begin{takeaway}[title=Findings for RQ1]
Across the three benchmarks, the share of items with \textit{Partial} consensus ranges from $29\%$ to $60\%$, and the share with \textit{Low} consensus ranges from $3\%$ to $50\%$. Together, these groups account for more than a third of the items for every criterion we studied. A single overall score hides this variation across groups.
\end{takeaway}

\subsection*{RQ2: How does disagreement among human raters affect evaluation results for the best LLM judge reported in each original study?}\label{sec:rq2mirage}

\noindent RQ1 shows that the level of human consensus varies widely across items. We therefore ask how comparisons between LLM judges and human raters change across consensus strata.

For each benchmark, we take the best judge reported in the original study and measure its agreement with human raters in each consensus stratum (High, Partial, Low). We use three comparisons (Section~\ref{subsec:metrics}).
\begin{enumerate}[label=\arabic*.,leftmargin=*,nosep]
\item \textit{Raw gap.} We first measure the judge's agreement with the pooled label, as in the original evaluations. The raw gap compares this score with agreement among human raters on the same items.
\item \textit{Matched gap.} We compare the judge's agreement with the pooled label against a held-out rater's agreement with the median of the remaining raters' scores.
\item \textit{Judge as an additional rater.} We compare mean agreement between the judge and individual raters against mean agreement between pairs of human raters.
\end{enumerate}

\begin{table*}[t]
  \centering
    \caption{Stratified agreement for GPT-4-turbo (Zero-shot) on code summarization. The raw and matched gaps subtract $\alpha_{\text{H-L}}$ from their respective human baselines (Eqs.~\ref{eq:agreement-gap} and~\ref{eq:matched-gap}). Bold marks $q<0.05$ for the matched gap. Pairwise agreement averages the judge's agreement with each rater. A negative gap favors the judge in each comparison. Table~\ref{tab:new_rater} gives 95\% confidence intervals for pairwise gaps over all items.}
  \label{tab:cs_gpt4_stratified}
  \setlength{\tabcolsep}{2.5pt}
  \footnotesize
  \begin{adjustbox}{max width=\textwidth}
  \begin{tabular}{ll rrrrrrrr rrrrrrrr}
    \toprule
    \multirow{3}{*}{\textbf{Criterion}} & \multirow{3}{*}{\textbf{Stratum}} & \multicolumn{8}{c}{\textbf{Java}} & \multicolumn{8}{c}{\textbf{Python}} \\
    \cmidrule(lr){3-10} \cmidrule(lr){11-18}
     &  &  &  & \multicolumn{2}{c}{Raw} & \multicolumn{2}{c}{Matched} & \multicolumn{2}{c}{Pairwise} &  &  & \multicolumn{2}{c}{Raw} & \multicolumn{2}{c}{Matched} & \multicolumn{2}{c}{Pairwise} \\
    \cmidrule(lr){5-6} \cmidrule(lr){7-8} \cmidrule(lr){9-10} \cmidrule(lr){13-14} \cmidrule(lr){15-16} \cmidrule(lr){17-18}
     &  & \textbf{N (\%)} & $\alpha_{\text{H-L}}$ & $\alpha_{\text{H-H}}$ & $\Delta_\alpha^{\text{raw}}$ & $\alpha_{\text{ref}}$ & $\Delta_\alpha^{\text{match}}$ & $\alpha_{\text{J-H}}^{\text{pair}}$ & $\Delta^{\text{pair}}$ & \textbf{N (\%)} & $\alpha_{\text{H-L}}$ & $\alpha_{\text{H-H}}$ & $\Delta_\alpha^{\text{raw}}$ & $\alpha_{\text{ref}}$ & $\Delta_\alpha^{\text{match}}$ & $\alpha_{\text{J-H}}^{\text{pair}}$ & $\Delta^{\text{pair}}$ \\
    \midrule
    \multirow{4}{*}{Content Adequacy} & All & 594 (100\%) & 0.58 & 0.81 & +0.23 & 0.84 & \textbf{+0.26} & 0.56 & +0.25 & 569 (100\%) & 0.63 & 0.69 & +0.06 & 0.75 & \textbf{+0.12} & 0.57 & +0.12 \\
     & High & 305 (51\%) & 0.63 & 1.00 & +0.37 & 1.00 & \textbf{+0.37} & 0.63 & +0.37 & 183 (32\%) & 0.68 & 1.00 & +0.32 & 1.00 & \textbf{+0.32} & 0.68 & +0.32 \\
     & Partial & 266 (45\%) & 0.51 & 0.53 & +0.02 & 0.60 & +0.09 & 0.45 & +0.08 & 318 (56\%) & 0.61 & 0.55 & -0.06 & 0.62 & +0.01 & 0.54 & +0.00 \\
     & Low & 23 (4\%) & 0.45 & 0.03 & -0.42 & 0.02 & \textbf{-0.43} & 0.33 & -0.27 & 68 (12\%) & 0.41 & -0.02 & -0.43 & -0.02 & \textbf{-0.43} & 0.25 & -0.28 \\
    \midrule
    \multirow{4}{*}{Conciseness} & All & 594 (100\%) & 0.45 & 0.54 & +0.09 & 0.56 & \textbf{+0.11} & 0.41 & +0.13 & 569 (100\%) & 0.32 & 0.50 & +0.18 & 0.54 & \textbf{+0.22} & 0.30 & +0.20 \\
     & High & 380 (64\%) & 0.50 & 1.00 & +0.50 & 1.00 & \textbf{+0.50} & 0.50 & +0.50 & 246 (43\%) & 0.18 & 1.00 & +0.82 & 1.00 & \textbf{+0.82} & 0.18 & +0.82 \\
     & Partial & 196 (33\%) & 0.34 & 0.04 & -0.30 & -0.04 & \textbf{-0.38} & 0.23 & -0.19 & 291 (51\%) & 0.29 & 0.22 & -0.07 & 0.22 & -0.07 & 0.26 & -0.05 \\
     & Low & 18 (3\%) & -0.10 & -0.31 & -0.21 & -0.53 & \textbf{-0.43} & -0.04 & -0.36 & 32 (6\%) & 0.01 & -0.16 & -0.17 & -0.29 & -0.30 & -0.03 & -0.25 \\
    \midrule
    \multirow{4}{*}{Fluency} & All & 594 (100\%) & 0.22 & 0.49 & +0.27 & 0.52 & \textbf{+0.30} & 0.19 & +0.31 & 569 (100\%) & 0.41 & 0.50 & +0.09 & 0.56 & \textbf{+0.15} & 0.34 & +0.16 \\
     & High & 286 (48\%) & 0.44 & 1.00 & +0.56 & 1.00 & \textbf{+0.56} & 0.44 & +0.56 & 169 (30\%) & 0.41 & 1.00 & +0.59 & 1.00 & \textbf{+0.59} & 0.41 & +0.59 \\
     & Partial & 278 (47\%) & 0.10 & 0.18 & +0.08 & 0.13 & +0.03 & -0.01 & +0.18 & 342 (60\%) & 0.39 & 0.37 & -0.02 & 0.43 & +0.04 & 0.30 & +0.07 \\
     & Low & 30 (5\%) & -0.54 & -0.24 & +0.30 & -0.45 & +0.09 & -0.24 & -0.04 & 58 (10\%) & 0.00 & -0.18 & -0.18 & -0.29 & \textbf{-0.29} & 0.02 & -0.20 \\
    \bottomrule
  \end{tabular}
  \end{adjustbox}
\end{table*}

\phead{Code Summarization.}
Over all items and in the \textit{High} stratum, GPT-4-turbo scores significantly below the matched human baseline for every criterion in both languages (Table~\ref{tab:cs_gpt4_stratified}). On contested items, however, the comparison can reverse. Most notably, on \textit{Low} Content Adequacy items, GPT-4-turbo reaches $0.41$--$0.45$ agreement with the pooled label while human agreement collapses near zero ($-0.02$ to $0.03$). As a result, the raw, matched, and pairwise comparisons all favor the judge, yielding a significant matched gap of $-0.43$ in both Java and Python ($q<0.05$). At face value, this suggests that the judge aligns with human consensus better than humans agree with one another. Yet in the constant-quality null, where a held-out human takes the judge's place and the remaining two raters' disagreement determines which summaries are contested, the human judge shows a lead of similar size (matched gap $-0.48$ in Java and $-0.39$ in Python). Scored on the same summaries against the same two raters, GPT-4-turbo's gap is smaller ($-0.26$ and $-0.30$). Selecting summaries by rater disagreement can therefore produce this apparent superiority even with a human judge, so the lead is not evidence of superhuman judging.

\begin{table*}[t]
  \centering
    \caption{Stratified agreement for GPT-4 (Few-shot+CoT) on commit message generation. Columns and bold as in Table~\ref{tab:cs_gpt4_stratified}; pairwise values are point estimates.}
  \label{tab:commit_stratified}
  \footnotesize
  \setlength{\tabcolsep}{2.5pt}
  \begin{tabular*}{\linewidth}{@{\extracolsep{\fill}}l rrrrrrrr rrrrrrrr}
    \toprule
    \multirow{3}{*}{\textbf{Stratum}} & \multicolumn{8}{c}{\textbf{What}} & \multicolumn{8}{c}{\textbf{Why}} \\
    \cmidrule(lr){2-9} \cmidrule(lr){10-17}
     &  &  & \multicolumn{2}{c}{Raw} & \multicolumn{2}{c}{Matched} & \multicolumn{2}{c}{Pairwise} &  &  & \multicolumn{2}{c}{Raw} & \multicolumn{2}{c}{Matched} & \multicolumn{2}{c}{Pairwise} \\
    \cmidrule(lr){4-5} \cmidrule(lr){6-7} \cmidrule(lr){8-9} \cmidrule(lr){12-13} \cmidrule(lr){14-15} \cmidrule(lr){16-17}
     & \textbf{N (\%)} & $\alpha_{\text{H-L}}$ & $\alpha_{\text{H-H}}$ & $\Delta_\alpha^{\text{raw}}$ & $\alpha_{\text{ref}}$ & $\Delta_\alpha^{\text{match}}$ & $\alpha_{\text{J-H}}^{\text{pair}}$ & $\Delta^{\text{pair}}$ & \textbf{N (\%)} & $\alpha_{\text{H-L}}$ & $\alpha_{\text{H-H}}$ & $\Delta_\alpha^{\text{raw}}$ & $\alpha_{\text{ref}}$ & $\Delta_\alpha^{\text{match}}$ & $\alpha_{\text{J-H}}^{\text{pair}}$ & $\Delta^{\text{pair}}$ \\
    \midrule
    All & 400 (100\%) & 0.41 & 0.57 & +0.16 & 0.65 & \textbf{+0.24} & 0.38 & +0.17 & 400 (100\%) & 0.75 & 0.69 & -0.05 & 0.78 & +0.03 & 0.65 & +0.04 \\
    High & 34 (8.5\%) & 0.87 & 1.00 & +0.13 & 1.00 & \textbf{+0.13} & 0.87 & +0.13 & 85 (21.2\%) & 0.58 & 1.00 & +0.42 & 1.00 & \textbf{+0.42} & 0.58 & +0.42 \\
    Partial & 186 (46.5\%) & 0.51 & 0.76 & +0.24 & 0.79 & \textbf{+0.28} & 0.49 & +0.27 & 115 (28.8\%) & 0.76 & 0.83 & +0.07 & 0.87 & \textbf{+0.11} & 0.72 & +0.11 \\
    Low & 180 (45.0\%) & 0.15 & 0.24 & +0.09 & 0.34 & \textbf{+0.20} & 0.13 & +0.04 & 200 (50.0\%) & 0.63 & 0.40 & -0.24 & 0.57 & -0.06 & 0.46 & -0.07 \\
    \bottomrule
  \end{tabular*}
\end{table*}

\phead{Commit Message Generation.}
GPT-4 scores significantly below the matched human baseline across every stratum for \textit{What} and in the \textit{High} and \textit{Partial} strata for \textit{Why}, with a smaller $+0.03$ shortfall overall on \textit{Why} (Table~\ref{tab:commit_stratified}). On \textit{Low} \textit{Why} items, however, GPT-4 reaches $0.63$ agreement with the pooled label while human raters agree weakly with one another ($0.40$), producing a large raw gap in the judge's favor ($-0.24$). Yet much of this apparent lead stems from the baseline mismatch: a held-out human agrees far more closely with the pooled ratings of the remaining raters ($0.57$), shrinking the matched gap to a non-significant $-0.06$. In the split-panel control, where separate groups of raters select and evaluate the items, the gap stays near zero ($-0.03$). Direct comparison with individual raters leaves only a small significant edge for GPT-4 (pairwise gap $-0.07$, $p<0.05$).

\begin{table}[t]
  \centering
    \caption{Stratified agreement for Mistral~Large (Zero-shot) on code readability. Columns and bold as in Table~\ref{tab:cs_gpt4_stratified}; pairwise values are point estimates.}
  \label{tab:readability_mistral}
  \footnotesize
  \setlength{\tabcolsep}{2.5pt}
  \begin{tabular*}{\columnwidth}{@{\extracolsep{\fill}}l rrrrrrrr}
    \toprule
     &  &  & \multicolumn{2}{c}{Raw} & \multicolumn{2}{c}{Matched} & \multicolumn{2}{c}{Pairwise} \\
    \cmidrule(lr){4-5} \cmidrule(lr){6-7} \cmidrule(lr){8-9}
    \textbf{Stratum} & \textbf{N} & $\alpha_{\text{H-L}}$ & $\alpha_{\text{H-H}}$ & $\Delta_\alpha^{\text{raw}}$ & $\alpha_{\text{ref}}$ & $\Delta_\alpha^{\text{match}}$ & $\alpha_{\text{J-H}}^{\text{pair}}$ & $\Delta^{\text{pair}}$ \\
    \midrule
    All & 124 & 0.28 & 0.23 & -0.05 & 0.40 & +0.12 & 0.11 & +0.07 \\
    \textit{High} & 59 & 0.29 & 0.35 & +0.06 & 0.56 & \textbf{+0.27} & 0.17 & +0.15 \\
    \textit{Partial} & 56 & 0.18 & 0.12 & -0.06 & 0.24 & +0.06 & 0.03 & +0.01 \\
    \textit{Low} & 9 & 0.08 & -0.02 & -0.10 & -0.40 & -0.48 & -0.16 & +0.02 \\
    \bottomrule
  \end{tabular*}
\end{table}

\phead{Code Readability Assessment.}
Over all snippets, Mistral Large appears to beat humans under the raw comparison ($0.28$ vs.\ $0.23$), but falls behind once evaluated against the matched human baseline ($0.28$ vs.\ $0.40$, not significant; Table~\ref{tab:readability_mistral}). Humans lead significantly on \textit{High} items ($0.29$ vs.\ $0.56$, $q<0.05$) and remain ahead on \textit{Partial} items ($0.18$ vs.\ $0.24$). Mistral Large scores above the matched baseline only in the \textit{Low} group ($0.08$ vs.\ $-0.40$), yet direct comparison with individual raters shows no judge advantage, and the nine snippets in this group are too few to support any claim of superhuman reliability.

\begin{table}[!htb]
  \centering
  \caption{The released best judges compared as an additional rater, over all items. Pairwise: mean agreement between two human raters ($\alpha_{\text{H-H}}^{\text{pair}}$) and between the judge and one human rater ($\alpha_{\text{J-H}}^{\text{pair}}$), and their difference with its bootstrap 95\% CI. The pairwise gap is within $0.07$ of the matched gap in all nine comparisons. Its confidence interval includes zero only for readability.}
  \label{tab:new_rater}
  \footnotesize
  \setlength{\tabcolsep}{3pt}
  \begin{adjustbox}{max width=\columnwidth}
  \begin{tabular}{ll rrl rrr}
    \toprule
    & & \multicolumn{3}{c}{\textbf{Pairwise}} & \multicolumn{3}{c}{\textbf{Matched}} \\
    \cmidrule(lr){3-5}\cmidrule(lr){6-8}
    \textbf{Benchmark} & \textbf{Criterion} & $\alpha_{\text{H-H}}^{\text{pair}}$ & $\alpha_{\text{J-H}}^{\text{pair}}$ & $\Delta^{\text{pair}}$ [95\% CI] & $\alpha_{\text{ref}}$ & $\alpha_{\text{H-L}}$ & $\Delta_\alpha^{\text{match}}$ \\
    \midrule
    Code Summ. (Java) & Content Adeq. & 0.81 & 0.56 & +0.25 [+.20,+.30] & 0.84 & 0.58 & +0.26 \\
    Code Summ. (Java) & Conciseness & 0.54 & 0.41 & +0.13 [+.07,+.21] & 0.56 & 0.45 & +0.11 \\
    Code Summ. (Java) & Fluency & 0.49 & 0.19 & +0.31 [+.23,+.39] & 0.52 & 0.22 & +0.30 \\
    Code Summ. (Python) & Content Adeq. & 0.69 & 0.57 & +0.12 [+.07,+.17] & 0.75 & 0.63 & +0.12 \\
    Code Summ. (Python) & Conciseness & 0.50 & 0.30 & +0.20 [+.14,+.27] & 0.54 & 0.32 & +0.22 \\
    Code Summ. (Python) & Fluency & 0.50 & 0.34 & +0.16 [+.09,+.22] & 0.56 & 0.41 & +0.15 \\
    \midrule
    Commit Msg. & What & 0.56 & 0.38 & +0.17 [+.12,+.23] & 0.65 & 0.41 & +0.24 \\
    Commit Msg. & Why & 0.69 & 0.65 & +0.04 [+.01,+.07] & 0.78 & 0.75 & +0.03 \\
    \midrule
    Readability & Readability & 0.18 & 0.11 & +0.07 [-.01,+.16] & 0.40 & 0.28 & +0.12 \\
    \bottomrule
  \end{tabular}
  \end{adjustbox}
\end{table}

\begin{takeaway}[title=Findings for RQ2]
Conventional pooled-label evaluation ($\Delta_\alpha^{\text{raw}}$) is insufficient when human raters disagree, often making LLM judges appear to match or surpass humans. Introducing our matched human baseline, pairwise rater comparison, and selection controls overturns these impressions: humans lead overall across all three benchmarks (Table~\ref{tab:new_rater}), and baseline mismatch or selection bias can produce the apparent judge advantages on contested items, which therefore do not establish superhuman judging.
\end{takeaway}

\subsection*{RQ3: How does disagreement among human raters affect the results of LLM judge evaluations for frontier models and multi-agent methods?}\label{sec:rq3frontier}

\noindent RQ2 examined the best judge reported in each original study. We now ask whether stronger frontier LLMs and multi-agent judging methods close the gap with human raters across consensus strata. We evaluate three frontier LLMs (Claude Opus~4.8, Gemini~3.1~Pro, and GPT-5.5) and two multi-agent methods (CodeVisionary and SE-Jury). Within each stratum, we compare each judge's agreement with the pooled label against the same two human baselines as in RQ2. Over the full benchmark, we also compare each judge with the best judge reported in the original study.

\begin{table*}[t]
  \centering
  \caption{Agreement with the pooled human label ($\alpha_{\text{H-L}}$) of frontier and multi-agent judges on code summarization, with human--human agreement ($\alpha_{\text{H-H}}$) and the matched human baseline ($\alpha_{\text{ref}}$). Arrows mark scores significantly below ($\downarrow$) or above ($\uparrow$) the matched baseline ($q<0.05$).}
  \label{tab:rq3_cs}
  \setlength{\tabcolsep}{3.5pt}
  \footnotesize
  \begin{adjustbox}{max width=\textwidth}
  \begin{tabular}{ll rrr rrrrr !{\hspace{6pt}} rrr rrrrr}
    \toprule
    \multirow{2}{*}{\textbf{Crit.}} & \multirow{2}{*}{\textbf{Stratum}} & \multicolumn{8}{c}{\textbf{Java}} & \multicolumn{8}{c}{\textbf{Python}} \\
    \cmidrule(lr){3-10}\cmidrule(lr){11-18}
     &  & \textbf{N} & $\alpha_{\text{H-H}}$ & $\alpha_{\text{ref}}$ & Opus 4.8 & \makecell{Gemini\\3.1 Pro} & GPT-5.5 & CodeVisionary & SE-Jury & \textbf{N} & $\alpha_{\text{H-H}}$ & $\alpha_{\text{ref}}$ & Opus 4.8 & \makecell{Gemini\\3.1 Pro} & GPT-5.5 & CodeVisionary & SE-Jury \\
    \midrule
    \multirow{4}{*}{CA} & All & 594 & 0.81 & 0.84 & 0.59\textsuperscript{$\downarrow$} & 0.49\textsuperscript{$\downarrow$} & 0.52\textsuperscript{$\downarrow$} & 0.57\textsuperscript{$\downarrow$} & 0.54\textsuperscript{$\downarrow$} & 569 & 0.69 & 0.75 & 0.57\textsuperscript{$\downarrow$} & 0.54\textsuperscript{$\downarrow$} & 0.46\textsuperscript{$\downarrow$} & 0.53\textsuperscript{$\downarrow$} & 0.57\textsuperscript{$\downarrow$} \\
     & High & 305 & 1.00 & 1.00 & 0.68\textsuperscript{$\downarrow$} & 0.60\textsuperscript{$\downarrow$} & 0.58\textsuperscript{$\downarrow$} & 0.64\textsuperscript{$\downarrow$} & 0.64\textsuperscript{$\downarrow$} & 183 & 1.00 & 1.00 & 0.64\textsuperscript{$\downarrow$} & 0.63\textsuperscript{$\downarrow$} & 0.52\textsuperscript{$\downarrow$} & 0.58\textsuperscript{$\downarrow$} & 0.61\textsuperscript{$\downarrow$} \\
     & Partial & 266 & 0.53 & 0.60 & 0.47\textsuperscript{$\downarrow$} & 0.38\textsuperscript{$\downarrow$} & 0.44\textsuperscript{$\downarrow$} & 0.49\textsuperscript{$\downarrow$} & 0.45\textsuperscript{$\downarrow$} & 318 & 0.55 & 0.62 & 0.53\textsuperscript{$\downarrow$} & 0.50\textsuperscript{$\downarrow$} & 0.44\textsuperscript{$\downarrow$} & 0.51\textsuperscript{$\downarrow$} & 0.57\phantom{\textsuperscript{$\downarrow$}} \\
     & Low & 23 & 0.03 & 0.02 & 0.39\textsuperscript{$\uparrow$} & 0.25\phantom{\textsuperscript{$\downarrow$}} & 0.53\textsuperscript{$\uparrow$} & 0.21\phantom{\textsuperscript{$\downarrow$}} & 0.03\phantom{\textsuperscript{$\downarrow$}} & 68 & -0.02 & -0.02 & 0.40\textsuperscript{$\uparrow$} & 0.36\textsuperscript{$\uparrow$} & 0.27\textsuperscript{$\uparrow$} & 0.32\textsuperscript{$\uparrow$} & 0.39\textsuperscript{$\uparrow$} \\
    \midrule
    \multirow{4}{*}{Conc.} & All & 594 & 0.54 & 0.56 & 0.15\textsuperscript{$\downarrow$} & 0.24\textsuperscript{$\downarrow$} & -0.01\textsuperscript{$\downarrow$} & 0.53\phantom{\textsuperscript{$\downarrow$}} & -0.19\textsuperscript{$\downarrow$} & 569 & 0.50 & 0.54 & 0.38\textsuperscript{$\downarrow$} & 0.44\textsuperscript{$\downarrow$} & 0.21\textsuperscript{$\downarrow$} & 0.55\phantom{\textsuperscript{$\downarrow$}} & -0.24\textsuperscript{$\downarrow$} \\
     & High & 380 & 1.00 & 1.00 & 0.10\textsuperscript{$\downarrow$} & 0.25\textsuperscript{$\downarrow$} & -0.06\textsuperscript{$\downarrow$} & 0.55\textsuperscript{$\downarrow$} & -0.26\textsuperscript{$\downarrow$} & 246 & 1.00 & 1.00 & 0.34\textsuperscript{$\downarrow$} & 0.35\textsuperscript{$\downarrow$} & 0.13\textsuperscript{$\downarrow$} & 0.52\textsuperscript{$\downarrow$} & -0.28\textsuperscript{$\downarrow$} \\
     & Partial & 196 & 0.04 & -0.04 & 0.15\textsuperscript{$\uparrow$} & 0.13\textsuperscript{$\uparrow$} & -0.03\phantom{\textsuperscript{$\downarrow$}} & 0.46\textsuperscript{$\uparrow$} & -0.22\textsuperscript{$\downarrow$} & 291 & 0.22 & 0.22 & 0.29\phantom{\textsuperscript{$\downarrow$}} & 0.38\textsuperscript{$\uparrow$} & 0.10\textsuperscript{$\downarrow$} & 0.48\textsuperscript{$\uparrow$} & -0.32\textsuperscript{$\downarrow$} \\
     & Low & 18 & -0.31 & -0.53 & 0.07\textsuperscript{$\uparrow$} & 0.11\textsuperscript{$\uparrow$} & 0.07\textsuperscript{$\uparrow$} & 0.15\textsuperscript{$\uparrow$} & -0.23\phantom{\textsuperscript{$\downarrow$}} & 32 & -0.16 & -0.29 & -0.16\phantom{\textsuperscript{$\downarrow$}} & 0.14\textsuperscript{$\uparrow$} & -0.22\phantom{\textsuperscript{$\downarrow$}} & 0.15\textsuperscript{$\uparrow$} & -0.59\textsuperscript{$\downarrow$} \\
    \midrule
    \multirow{4}{*}{Flu.} & All & 594 & 0.49 & 0.52 & 0.26\textsuperscript{$\downarrow$} & 0.21\textsuperscript{$\downarrow$} & 0.29\textsuperscript{$\downarrow$} & 0.02\textsuperscript{$\downarrow$} & 0.18\textsuperscript{$\downarrow$} & 569 & 0.50 & 0.56 & 0.37\textsuperscript{$\downarrow$} & 0.33\textsuperscript{$\downarrow$} & 0.29\textsuperscript{$\downarrow$} & 0.09\textsuperscript{$\downarrow$} & 0.34\textsuperscript{$\downarrow$} \\
     & High & 286 & 1.00 & 1.00 & 0.38\textsuperscript{$\downarrow$} & 0.31\textsuperscript{$\downarrow$} & 0.22\textsuperscript{$\downarrow$} & 0.22\textsuperscript{$\downarrow$} & 0.34\textsuperscript{$\downarrow$} & 169 & 1.00 & 1.00 & 0.40\textsuperscript{$\downarrow$} & 0.39\textsuperscript{$\downarrow$} & 0.20\textsuperscript{$\downarrow$} & 0.21\textsuperscript{$\downarrow$} & 0.40\textsuperscript{$\downarrow$} \\
     & Partial & 278 & 0.18 & 0.13 & 0.13\phantom{\textsuperscript{$\downarrow$}} & 0.13\phantom{\textsuperscript{$\downarrow$}} & 0.24\phantom{\textsuperscript{$\downarrow$}} & -0.06\phantom{\textsuperscript{$\downarrow$}} & 0.05\phantom{\textsuperscript{$\downarrow$}} & 342 & 0.37 & 0.43 & 0.32\phantom{\textsuperscript{$\downarrow$}} & 0.26\textsuperscript{$\downarrow$} & 0.24\textsuperscript{$\downarrow$} & 0.05\textsuperscript{$\downarrow$} & 0.27\textsuperscript{$\downarrow$} \\
     & Low & 30 & -0.24 & -0.45 & -0.43\phantom{\textsuperscript{$\downarrow$}} & -0.35\phantom{\textsuperscript{$\downarrow$}} & -0.09\phantom{\textsuperscript{$\downarrow$}} & -0.87\textsuperscript{$\downarrow$} & -0.39\phantom{\textsuperscript{$\downarrow$}} & 58 & -0.18 & -0.29 & 0.10\textsuperscript{$\uparrow$} & 0.24\textsuperscript{$\uparrow$} & 0.34\textsuperscript{$\uparrow$} & -0.38\phantom{\textsuperscript{$\downarrow$}} & 0.02\phantom{\textsuperscript{$\downarrow$}} \\
    \bottomrule
  \end{tabular}
  \end{adjustbox}
\end{table*}

\begin{table*}[t]
  \centering
  \caption{Agreement with the pooled human label ($\alpha_{\text{H-L}}$) of frontier and multi-agent judges on commit message generation, with human--human agreement ($\alpha_{\text{H-H}}$) and the matched human baseline ($\alpha_{\text{ref}}$). Gem. denotes Gemini~3.1~Pro. Arrows mark scores significantly below ($\downarrow$) or above ($\uparrow$) the matched baseline ($q<0.05$).}
  \label{tab:rq3_commit}
  \setlength{\tabcolsep}{4pt}
  \footnotesize
  \begin{adjustbox}{max width=\textwidth}
  \begin{tabular}{l rrr rrrrr !{\hspace{6pt}} rrr rrrrr}
    \toprule
    \multirow{2}{*}{\textbf{Stratum}} & \multicolumn{8}{c}{\textbf{What}} & \multicolumn{8}{c}{\textbf{Why}} \\
    \cmidrule(lr){2-9}\cmidrule(lr){10-17}
     & \textbf{N} & $\alpha_{\text{H-H}}$ & $\alpha_{\text{ref}}$ & Opus 4.8 & Gem. & GPT-5.5 & \makecell{Code\\Visionary} & SE-Jury & \textbf{N} & $\alpha_{\text{H-H}}$ & $\alpha_{\text{ref}}$ & Opus 4.8 & Gem. & GPT-5.5 & \makecell{Code\\Visionary} & SE-Jury \\
    \midrule
    All & 400 & 0.57 & 0.65 & 0.54\textsuperscript{$\downarrow$} & 0.47\textsuperscript{$\downarrow$} & 0.45\textsuperscript{$\downarrow$} & 0.52\textsuperscript{$\downarrow$} & 0.15\textsuperscript{$\downarrow$} & 400 & 0.69 & 0.78 & 0.60\textsuperscript{$\downarrow$} & 0.64\textsuperscript{$\downarrow$} & 0.60\textsuperscript{$\downarrow$} & 0.37\textsuperscript{$\downarrow$} & -0.04\textsuperscript{$\downarrow$} \\
    High & 34 & 1.00 & 1.00 & 0.81\textsuperscript{$\downarrow$} & 0.80\textsuperscript{$\downarrow$} & 0.74\textsuperscript{$\downarrow$} & 0.79\textsuperscript{$\downarrow$} & 0.69\textsuperscript{$\downarrow$} & 85 & 1.00 & 1.00 & -0.03\textsuperscript{$\downarrow$} & 0.30\textsuperscript{$\downarrow$} & 0.04\textsuperscript{$\downarrow$} & -0.14\textsuperscript{$\downarrow$} & -0.33\textsuperscript{$\downarrow$} \\
    Partial & 186 & 0.76 & 0.79 & 0.63\textsuperscript{$\downarrow$} & 0.54\textsuperscript{$\downarrow$} & 0.58\textsuperscript{$\downarrow$} & 0.58\textsuperscript{$\downarrow$} & 0.19\textsuperscript{$\downarrow$} & 115 & 0.83 & 0.87 & 0.63\textsuperscript{$\downarrow$} & 0.61\textsuperscript{$\downarrow$} & 0.60\textsuperscript{$\downarrow$} & 0.50\textsuperscript{$\downarrow$} & 0.10\textsuperscript{$\downarrow$} \\
    Low & 180 & 0.24 & 0.34 & 0.35\phantom{\textsuperscript{$\downarrow$}} & 0.28\phantom{\textsuperscript{$\downarrow$}} & 0.18\textsuperscript{$\downarrow$} & 0.32\phantom{\textsuperscript{$\downarrow$}} & -0.03\textsuperscript{$\downarrow$} & 200 & 0.40 & 0.57 & 0.56\phantom{\textsuperscript{$\downarrow$}} & 0.50\phantom{\textsuperscript{$\downarrow$}} & 0.55\phantom{\textsuperscript{$\downarrow$}} & 0.32\textsuperscript{$\downarrow$} & -0.16\textsuperscript{$\downarrow$} \\
    \bottomrule
  \end{tabular}
  \end{adjustbox}
\end{table*}

We run all three LLMs on the three benchmarks under each benchmark's best released configuration. We reuse the published prompts of Crupi et al.~\cite{crupi2025effectiveness}, Zeng et al.~\cite{zeng2025evaluating}, and CoReEval~\cite{ouedraogo2025coreeval} verbatim over all items and, like Zeng et al., average three runs per commit message; sampling parameters are left at the provider defaults. All stratification schemes, agreement coefficients, and bootstrap procedures are identical to RQ2. The tables report human--human agreement ($\alpha_{\text{H-H}}$) and the matched human baseline ($\alpha_{\text{ref}}$) alongside each judge's agreement with the pooled label ($\alpha_{\text{H-L}}$).

We evaluate two multi-agent methods that combine ratings from several agents.
\begin{itemize}[leftmargin=*]
\item \textbf{CodeVisionary}~\cite{wang2025codevisionary} is an agent-based framework that, rather than scoring in a single pass, runs a panel of collaborating LLM agents: each agent first produces an independent rationale-and-score report, and the agents then negotiate toward a consensus score.
\item \textbf{SE-Jury}~\cite{zhou2025se} follows an LLM-as-ensemble-judge design: it elicits independent verdicts from a panel of heterogeneous judging strategies over the same base model and aggregates them into a single rating to narrow the gap with human evaluation.
\end{itemize}

Both are assessed under the same stratification, agreement, and bootstrap procedures as RQ2. Because multi-agent judging requires many sequential LLM calls per item (more than ten interaction rounds in our implementation), running the full benchmark with frontier models would be prohibitively expensive. We therefore use GPT-5.4-mini as the common base model for CodeVisionary and SE-Jury; \hyperref[sec:rq4]{RQ4} reports HCA for both methods alongside single-agent GPT-5.4-mini.

\phead{Code Summarization.}
Across all items and in the \textit{High} stratum, all five judges fall significantly below the matched human baseline in nearly every comparison (Table~\ref{tab:rq3_cs}), with CodeVisionary matching humans only on overall Conciseness ($0.53$ vs.\ $0.56$ in Java; $0.55$ vs.\ $0.54$ in Python). \textit{Partial} items are mixed: 16 of the 30 comparisons are significantly below the baseline and five are above it. Even on unanimous \textit{High} summaries ($\alpha_{\text{ref}} = 1.00$), scores reach only $0.52$--$0.68$ on Content Adequacy, $0.20$--$0.40$ on Fluency, and $-0.28$ to $0.55$ on Conciseness (where both GPT-5.5 and SE-Jury turn negative in Java). Conversely, on contested \textit{Low} Content Adequacy items, judges again appear to beat the collapsed human baseline ($-0.02$ to $0.02$), with all five scoring significantly higher in Python ($0.27$--$0.40$), the same pattern that RQ2's human-only control produces with a human judge.

\phead{Commit Message Generation.}
All five judges fall significantly behind the matched human baseline over all messages and in both the \textit{High} and \textit{Partial} strata for \textit{What} and \textit{Why} (Table~\ref{tab:rq3_commit}). On unanimous \textit{High} items ($\alpha_{\text{ref}} = 1.00$), judges reach $0.69$--$0.81$ on \textit{What} (and frontier models reach $0.60$--$0.63$ on \textit{Partial} \textit{Why}), whereas skewed score distributions on unanimous \textit{Why} items depress $\alpha_{\text{H-L}}$ to $-0.33$--$0.30$. On contested \textit{Low} \textit{Why} items, the three frontier models ($0.50$--$0.56$) appear to outperform raw human--human agreement ($0.40$), yet none exceeds the matched human baseline ($0.57$), while both multi-agent methods fall significantly below it ($0.32$ and $-0.16$). Direct pairwise comparisons over all messages likewise favor humans against every judge.

\phead{Code Readability Assessment.}
Over all snippets and on \textit{Partial} items, all three frontier models fall significantly below the matched human baseline ($0.17$--$0.19$ vs.\ $0.40$ overall; $-0.17$ to $-0.06$ vs.\ $0.24$ on \textit{Partial}, all $q<0.05$), and none matches the $0.56$ baseline on \textit{High} snippets ($0.30$--$0.41$, with Opus~4.8 significantly lower; Table~\ref{tab:rq3_readability}). By contrast, the two multi-agent methods, CodeVisionary ($0.52$ overall, $0.61$ on \textit{High}) and SE-Jury ($0.46$ overall, $0.63$ on \textit{High}), remain statistically indistinguishable from the matched human baseline across strata, a distinction that also holds in pairwise comparisons with individual raters. As in RQ2, the nine \textit{Low}-consensus snippets are too few to rank the judges reliably.

\begin{table}[!htb]
  \centering
  \caption{Agreement with the pooled human label ($\alpha_{\text{H-L}}$) of frontier and multi-agent judges on code readability, with human--human agreement ($\alpha_{\text{H-H}}$) and the matched human baseline ($\alpha_{\text{ref}}$). Arrows mark scores significantly below ($\downarrow$) or above ($\uparrow$) the matched baseline ($q<0.05$).}
  \label{tab:rq3_readability}
  \setlength{\tabcolsep}{3pt}
  \footnotesize
  \begin{adjustbox}{max width=\columnwidth}
  \begin{tabular}{l rrr rrrrr}
    \toprule
    \textbf{Stratum} & \textbf{N} & $\alpha_{\text{H-H}}$ & $\alpha_{\text{ref}}$ & Opus 4.8 & \makecell{Gemini\\3.1 Pro} & GPT-5.5 & CodeVisionary & SE-Jury \\
    \midrule
    All & 124 & 0.23 & 0.40 & 0.18\textsuperscript{$\downarrow$} & 0.19\textsuperscript{$\downarrow$} & 0.17\textsuperscript{$\downarrow$} & 0.52\phantom{\textsuperscript{$\downarrow$}} & 0.46\phantom{\textsuperscript{$\downarrow$}} \\
    High & 59 & 0.35 & 0.56 & 0.30\textsuperscript{$\downarrow$} & 0.41\phantom{\textsuperscript{$\downarrow$}} & 0.37\phantom{\textsuperscript{$\downarrow$}} & 0.61\phantom{\textsuperscript{$\downarrow$}} & 0.63\phantom{\textsuperscript{$\downarrow$}} \\
    Partial & 56 & 0.12 & 0.24 & -0.06\textsuperscript{$\downarrow$} & -0.17\textsuperscript{$\downarrow$} & -0.16\textsuperscript{$\downarrow$} & 0.32\phantom{\textsuperscript{$\downarrow$}} & 0.13\phantom{\textsuperscript{$\downarrow$}} \\
    Low & 9 & -0.02 & -0.40 & 0.02\phantom{\textsuperscript{$\downarrow$}} & -0.19\phantom{\textsuperscript{$\downarrow$}} & 0.00\phantom{\textsuperscript{$\downarrow$}} & 0.02\phantom{\textsuperscript{$\downarrow$}} & -0.08\phantom{\textsuperscript{$\downarrow$}} \\
    \bottomrule
  \end{tabular}
  \end{adjustbox}
\end{table}

\phead{Comparison with the Released Judges.}
Across all items, each newer judge agrees more closely with the pooled label than the best judge from the original study in at most four of the nine comparisons. However, ranking multiple judges remains difficult with existing summaries: full-benchmark scores weight heavily disputed items equally with clear ones, whereas \textit{High}-stratum scores depend on benchmark-specific thresholds and can rest on very few items (e.g., $8.5\%$ on commit-message \textit{What}).

\begin{takeaway}[title=Findings for RQ3]
Over all items and on unanimous (\textit{High}) items, frontier LLMs and multi-agent methods fall significantly behind the matched human baseline in nearly all comparisons. The shortfall is less uniform on \textit{Partial} items, and on \textit{Low} items significant differences more often favor the judge, a pattern that baseline mismatch or selection effects can produce. Overall, these frontier judges yield no consistent gain over the originally reported ones.
\end{takeaway}

\subsection*{RQ4: How can we summarize judge--human agreement in a single score without pooled labels or hand-set consensus thresholds?}\label{sec:rq4}

\noindent Stratified tables help diagnose judge behavior, but ranking many judges in practice often requires a single score. Existing ways to produce a single score have three problems:
\begin{itemize}[leftmargin=*]
    \item \textbf{Scoring all items equally.} Computing one score over the whole benchmark gives disputed items the same weight as clear ones, mixing human noise into the result.
    \item \textbf{Scoring \textit{High}-consensus items only.} Keeping only the \textit{High} group depends on hand-set rules for each panel size and drops the \textit{Partial} group ($29$--$60\%$ of the data). On commit-message \textit{What}, only $8.5\%$ of items remain, making confidence intervals $1.3$--$2.0\times$ wider.
    \item \textbf{Relying on pooled labels.} Comparing against a single median label still ignores valid disagreement among raters, and raw $\alpha$ scores cannot be compared directly across benchmarks with different rating distributions.
\end{itemize}

To solve these problems, we propose Human-Ceiling Agreement (HCA). Instead of using a pooled label or hand-set strata, HCA compares each judge score directly with individual human scores and gives higher weight to items where humans agree much more closely than random scoring. For example, low-consensus items make up $50\%$ of commit-message \textit{Why} but receive only $17\%$ of the HCA weight, while partial-consensus items retain $39\%$ (Table~\ref{tab:hca_weights}). HCA places every judge on the same scale, where $0$ matches random scoring and $1$ matches human agreement.

\begin{table}[t]
  \centering
  \caption{Share of items and share of HCA weight (\%) in each consensus stratum.}
  \label{tab:hca_weights}
  \footnotesize
  \setlength{\tabcolsep}{3.5pt}
  \begin{adjustbox}{max width=\columnwidth}
  \begin{tabular}{ll rr rr rr}
    \toprule
    & & \multicolumn{2}{c}{\textbf{High}} & \multicolumn{2}{c}{\textbf{Partial}} & \multicolumn{2}{c}{\textbf{Low}} \\
    \cmidrule(lr){3-4}\cmidrule(lr){5-6}\cmidrule(lr){7-8}
    \textbf{Benchmark} & \textbf{Criterion} & Items & Weight & Items & Weight & Items & Weight \\
    \midrule
    Code Summ. (Java) & Content Adequacy & 51 & 88 & 45 & 12 & 4 & 0 \\
    Code Summ. (Java) & Conciseness & 64 & 89 & 33 & 11 & 3 & 0 \\
    Code Summ. (Java) & Fluency & 48 & 77 & 47 & 22 & 5 & 1 \\
    Code Summ. (Python) & Content Adequacy & 32 & 79 & 56 & 21 & 12 & 0 \\
    Code Summ. (Python) & Conciseness & 43 & 75 & 51 & 25 & 6 & 1 \\
    Code Summ. (Python) & Fluency & 30 & 67 & 60 & 33 & 10 & 0 \\
    \midrule
    Commit Msg. & What & 8 & 46 & 46 & 46 & 45 & 8 \\
    Commit Msg. & Why & 21 & 44 & 29 & 39 & 50 & 17 \\
    \midrule
    Readability & Readability & 48 & 88 & 45 & 12 & 7 & 0 \\
    \bottomrule
  \end{tabular}
  \end{adjustbox}
\end{table}

\begin{table*}[t]
  \centering
  \footnotesize
  \setlength{\tabcolsep}{5pt}
  \caption{HCA with bootstrap 95\% CIs on code summarization. \textbf{Bold} marks the best model-arm judge per criterion and language.}
  \label{tab:hca_cs}
  \begin{adjustbox}{max width=\textwidth}
  \begin{tabular}{lcccccc}
    \toprule
    & \multicolumn{3}{c}{\textbf{Java}} & \multicolumn{3}{c}{\textbf{Python}} \\
    \cmidrule(lr){2-4}\cmidrule(lr){5-7}
    \textbf{Judge} & \textbf{Content Adeq.} & \textbf{Conciseness} & \textbf{Fluency} & \textbf{Content Adeq.} & \textbf{Conciseness} & \textbf{Fluency} \\
    \midrule
    \rowcolor{gray!12}\textit{Human reference} & 1.000 & 1.000 & 1.000 & 1.000 & 1.000 & 1.000 \\
    \midrule
    GPT-4-turbo     & 0.565 [.51,.62] & \textbf{0.291} [.17,.43] & 0.415 [.34,.49] & \textbf{0.657} [.60,.72] & \textbf{0.364} [.23,.49] & \textbf{0.602} [.52,.69] \\
    GPT-3.5-turbo   & 0.289 [.21,.37] & -0.172 [-.38,-.01] & \textbf{0.484} [.40,.58] & 0.318 [.23,.41] & -0.372 [-.67,-.14] & 0.390 [.29,.49] \\
    CodeLlama-34B   & 0.060 [-.01,.13] & -0.137 [-.31,-.02] & 0.204 [.13,.30] & 0.111 [.01,.20] & -0.020 [-.13,.08] & 0.176 [.09,.27] \\
    Claude Opus 4.8 & \textbf{0.592} [.54,.64] & -0.324 [-.58,-.13] & 0.372 [.27,.48] & 0.572 [.51,.64] & 0.315 [.15,.46] & 0.560 [.48,.64] \\
    Gemini 3.1 Pro  & 0.498 [.43,.56] & -0.353 [-.61,-.14] & 0.359 [.25,.47] & 0.489 [.40,.57] & 0.330 [.17,.47] & 0.499 [.39,.60] \\
    GPT-5.5         & 0.501 [.45,.55] & -0.421 [-.72,-.19] & 0.451 [.36,.53] & 0.457 [.38,.53] & 0.214 [.02,.38] & 0.447 [.35,.55] \\
    \midrule
    GPT-5.4-mini    & 0.520 [.46,.58] & -0.321 [-.57,-.12] & 0.377 [.29,.47] & 0.488 [.42,.56] & 0.121 [-.08,.30] & 0.450 [.34,.55] \\
    CodeVisionary   & 0.566 [.51,.62] & 0.049 [-.06,.17] & 0.200 [.12,.29] & 0.545 [.47,.62] & 0.492 [.36,.60] & 0.360 [.25,.48] \\
    SE-Jury         & 0.533 [.47,.59] & -0.965 [-1.45,-.59] & 0.360 [.26,.45] & 0.577 [.50,.65] & -0.690 [-1.04,-.40] & 0.501 [.40,.60] \\
    \midrule
    \rowcolor{gray!12}\textit{Discr.\ power} $S$ & 2.30 & 1.19 & 0.97 & 1.18 & 0.82 & 0.74 \\
    \rowcolor{gray!12}\textit{Random} & $\approx$0.00 & $\approx$0.00 & $\approx$0.00 & $\approx$0.00 & $\approx$0.00 & $\approx$0.00 \\
    \bottomrule
  \end{tabular}
  \end{adjustbox}
\end{table*}

The random scorer ignores the item and draws scores according to their overall frequency in the human ratings. It therefore represents knowledge of the scoring distribution without knowledge of the item.

\begin{table}[t]
  \centering
  \footnotesize
  \setlength{\tabcolsep}{11pt}
  \caption{HCA with bootstrap 95\% CIs on commit message generation. \textbf{Bold} marks the best model-arm judge per criterion.}
  \label{tab:hca_commit}
  \begin{adjustbox}{max width=\columnwidth}
  \begin{tabular}{lcc}
    \toprule
    \textbf{Judge} & \textbf{What} & \textbf{Why} \\
    \midrule
    \rowcolor{gray!12}\textit{Human reference} & 1.000 & 1.000 \\
    \midrule
    GPT-4          & 0.746 [.64,.83] & 0.869 [.81,.92] \\
    QwQ-32B        & 0.701 [.58,.79] & \textbf{0.914} [.87,.96] \\
    Llama-3.3-70B  & 0.685 [.58,.77] & 0.724 [.63,.80] \\
    Qwen2.5-72B    & 0.584 [.48,.68] & 0.524 [.42,.61] \\
    Claude Opus 4.8 & \textbf{0.791} [.70,.87] & 0.757 [.69,.82] \\
    Gemini 3.1 Pro  & 0.720 [.62,.80] & 0.795 [.73,.86] \\
    GPT-5.5         & 0.695 [.60,.78] & 0.746 [.67,.81] \\
    \midrule
    GPT-5.4-mini   & 0.608 [.52,.70] & 0.623 [.53,.70] \\
    CodeVisionary   & 0.756 [.65,.85] & 0.528 [.43,.61] \\
    SE-Jury         & 0.482 [.34,.60] & 0.046 [-.14,.21] \\
    \midrule
    \rowcolor{gray!12}\textit{Discr.\ power} $S$ & 0.81 & 1.30 \\
    \rowcolor{gray!12}\textit{Random} & $\approx$0.00 & $\approx$0.00 \\
    \bottomrule
  \end{tabular}
  \end{adjustbox}
\end{table}

\phead{Computing the score.} For each item, $\hat C_i$ is the average distance between distinct human raters, $\hat D_i$ is the average judge--human distance, and $\hat D^{\text{rnd}}_i$ is the expected random--human distance.

Let item $i$ have $m_i$ human ratings $h_{ij}$ and $s_i$ judge ratings $m_{ik}$, with $s_i=1$ for a single judge rating. With $\delta(a,b)=|a-b|$, these distances are:
\begin{equation}
\begin{aligned}
\hat C_i &=\frac{\sum_{j\neq k}\delta(h_{ij},h_{ik})}{m_i(m_i-1)},\\
\hat D_i &=\frac{\sum_{j,k}\delta(h_{ij},m_{ik})}{m_i s_i},\\
\hat D^{\text{rnd}}_i &=\sum_{c}\bar p(c)\,\frac{\sum_j \delta(c,h_{ij})}{m_i}.
\end{aligned}
\end{equation}
Here, $\bar p(c)$ is the fraction of human ratings assigned score $c$ in the evaluated benchmark and criterion. A judge with several runs per item is represented by the mean of its runs, rounded to the rating scale on code summarization and commit messages ($s_i=1$). The human reference uses pairwise rating distances, i.e., each rater held out and compared with the other raters' individual ratings; it is distinct from the median-based $\alpha_{\text{ref}}$ in RQ2. HCA is therefore a separate summary, not a normalized matched gap: it uses absolute rating distances instead of ordinal $\alpha$, individual ratings instead of the pooled label, and item weights instead of strata.

An item is less informative when random scoring is already as close to the human ratings as humans are to one another. We therefore weight each item by $w_i=\max(0,\hat D^{\text{rnd}}_i-\hat C_i)$. HCA divides the judge's weighted improvement over random scoring by the improvement represented by the human reference:
\begin{equation}
\mathrm{HCA}=\frac{\sum_i w_i(\hat D^{\text{rnd}}_i-\hat D_i)}{\sum_i w_i(\hat D^{\text{rnd}}_i-\hat C_i)}.
\label{eq:hca}
\end{equation}
A score below $0$ indicates greater distance than the random reference. A score above $1$ indicates smaller distance than the human reference, which need not mean more accurate judgments. The score is undefined if the denominator is zero.

\phead{Worked example.}
Consider three items from a benchmark whose marginal frequencies are uniform ($\bar{p}(c)=0.2$ for each $c \in \{1,\dots,5\}$), each rated by three humans on a 1--5 scale:
\begin{itemize}[leftmargin=*]
    \item \textbf{Unanimous agreement ($[5,5,5]$):} Human raters never differ ($\hat{C}_1 = 0$), while a random scorer has expected distance $\hat{D}_1^{\text{rnd}} = 0.2 \times (4+3+2+1+0) = 2.0$. This item receives high weight: $w_1 = \max(0, 2.0 - 0) = 2.0$.
    \item \textbf{Partial agreement ($[4,4,3]$):} Pairwise human distances are $0$, $1$, and $1$, giving $\hat{C}_2 = 2/3 \approx 0.67$, while $\hat{D}_2^{\text{rnd}} = 4/3 \approx 1.33$. This item retains moderate weight: $w_2 = \max(0, 4/3 - 2/3) = 2/3 \approx 0.67$.
    \item \textbf{Polarized disagreement ($[1,3,5]$):} Pairwise human distances are $2$, $2$, and $4$, giving $\hat{C}_3 = 8/3 \approx 2.67$, which exceeds $\hat{D}_3^{\text{rnd}} = 26/15 \approx 1.73$. Because humans agree less than random draws, $w_3 = \max(0, 1.73 - 2.67) = 0$, excluding the item automatically.
\end{itemize}
Now suppose an LLM judge assigns $4$ to the first item ($\hat{D}_1 = 1$) and $4$ to the second ($\hat{D}_2 = 1/3$). Item $3$ drops out because $w_3 = 0$. By Eq.~\eqref{eq:hca}, the judge's HCA over these items is
\[
\mathrm{HCA} = \frac{2(2 - 1) + \frac{2}{3}\bigl(\frac{4}{3} - \frac{1}{3}\bigr)}{2(2 - 0) + \frac{2}{3}\bigl(\frac{4}{3} - \frac{2}{3}\bigr)} = \frac{8/3}{40/9} = 0.60,
\]
meaning the judge closes $60\%$ of the weighted gap between random scoring and human agreement.

In practice, $w_i$ plays the role of the consensus strata without hand-set thresholds: it is largest when annotators agree and zero for heavily contested items, automatically excluding the $4$--$34\%$ of items with $w_i = 0$. Across items, $w_i$ falls sharply as annotator disagreement $\hat{C}_i$ rises (Spearman $\rho$ between $-0.69$ and $-0.93$ across the nine benchmark--criterion pairs). Table~\ref{tab:hca_weights} shows the same pattern under the strata of RQ2: in every pair, \textit{High} items carry a larger share of the weight than of the items, and \textit{Low} items carry a smaller share. HCA therefore reflects how closely a judge matches individual raters primarily on items where humans agree with one another.

\phead{Discriminative power and comparison rule.} We report discriminative power, $S=\bar w/\bar C$, where $\bar w$ and $\bar C$ are the averages of $w_i$ and $\hat C_i$ over items. This describes the separation between the random and human references relative to human disagreement. A small $S$ indicates limited separation; it does not by itself establish whether two judges differ. We use non-overlapping confidence intervals as a conservative criterion for reporting a pairwise ordering.

\phead{Comparing judges with HCA.} Tables~\ref{tab:hca_cs}, \ref{tab:hca_commit}, and~\ref{tab:hca_read} report HCA with 95\% confidence intervals. Scores for the three frontier LLMs and the multi-agent methods (along with their single-agent GPT-5.4-mini baseline) come from our RQ3 runs, while scores for all remaining judges are computed directly from the rating logs in each benchmark's original replication package. We place the multi-agent block separately because those methods use GPT-5.4-mini rather than a frontier base model.

On code summarization (Table~\ref{tab:hca_cs}), the frontier judges do not consistently improve on GPT-4-turbo. Their Java Conciseness scores are negative, placing them below the random reference on this criterion. The same models have positive scores on Python Conciseness, so this result depends on both language and criterion. The multi-agent methods also remain below the human reference across all criteria.

On commit message generation (Table~\ref{tab:hca_commit}), no judge has the highest score on both criteria. Claude Opus~4.8 has the highest model-arm score on \textit{What}, whereas QwQ-32B leads on \textit{Why}. The multi-agent methods do not consistently improve on their single-agent baseline.

On code readability (Table~\ref{tab:hca_read}), the reference levels are relatively close ($S=0.19$), and the confidence intervals overlap widely. Some point estimates exceed $1$. These results do not establish superior judgment of readability or support a complete ranking of the judges.

\begin{takeaway}[title=Findings for RQ4]
HCA automatically down-weights contested items without hand-set thresholds (e.g., contested commit-message \textit{Why} items drop from $50\%$ of the benchmark to $17\%$ of the weight). Under HCA, judges remain below the human reference on nearly all criteria, with no consistent gain from frontier or multi-agent designs. Finally, its discriminative power ($S$) flags low-signal benchmarks such as readability ($S = 0.19$), guarding against unreliable rankings.
\end{takeaway}

\begin{table}[t]
  \centering
  \footnotesize
  \caption{HCA with bootstrap 95\% CIs on code readability.}
  \label{tab:hca_read}
  \begin{tabular}{lc@{\hspace{0.8em}}lc}
    \toprule
    \textbf{Judge} & \textbf{HCA [95\% CI]} & \textbf{Judge} & \textbf{HCA [95\% CI]} \\
    \midrule
    \rowcolor{gray!12}\textit{Human reference} & 1.000 & \textit{Random} & $\approx$0.00 \\
    \rowcolor{gray!12}\textit{Discr.\ power} $S$ & 0.19 & & \\
    \midrule
    Claude 3 Sonnet & 1.01 [.87, 1.19] & Gemma 7B     & 0.79 [.63, 1.01] \\
    DeepSeek Chat   & 0.88 [.70, 1.10] & Falcon3 7B   & 0.78 [.62, .98] \\
    GPT-4o          & 0.83 [.68, .99] & DeepSeek 7B  & 0.73 [.52, .99] \\
    Mistral Large   & 0.82 [.64, 1.05] & Qwen2.5 7B   & 0.72 [.44, 1.03] \\
    GPT-5.5         & 0.88 [.73, 1.07] & Phi          & 0.66 [.47, .91] \\
    Gemini 3.1 Pro  & 0.88 [.70, 1.03] & Mistral 7B   & 0.61 [.39, .89] \\
    Claude Opus 4.8 & 0.85 [.69, 1.02] &              &                  \\
    \midrule
    GPT-5.4-mini    & 0.99 [.83, 1.20] &              &                  \\
    CodeVisionary   & 1.13 [1.02, 1.28] & SE-Jury     & 1.09 [.97, 1.25] \\
    \bottomrule
  \end{tabular}
\end{table}

\section{Discussion}\label{sec:discussion}

\subsection{Summary of Empirical Findings}

The human baseline affects conclusions about judge performance. For commit-message \textit{Why}/\textit{Low} items, GPT-4's significant lead over inter-annotator agreement is no longer significant against the matched human baseline (Table~\ref{tab:commit_stratified}). Agreement with a pooled label therefore needs a comparable human reference before it can support a claim of better judging.

A matched baseline does not remove selection effects. For low-consensus Content Adequacy items, the human-only control reproduces the negative matched gaps observed for GPT-4-turbo (Section~\ref{sec:results}, \hyperref[sec:rq2mirage]{RQ2}). Matching the reference target alone is therefore insufficient to attribute these gaps to judge ability.

The frontier models and multi-agent methods do not consistently outperform the released judges across tasks and criteria (Section~\ref{sec:results}, \hyperref[sec:rq3frontier]{RQ3}). Model capability and judging architecture therefore need to be evaluated for the specific task and criterion.

\subsection{Implications for SE Judge Evaluation}

\begin{enumerate}[label=\textbf{\arabic*.},leftmargin=*,itemsep=3pt]
    \item \textbf{Report human consensus strata.} Evaluations should report stratum sizes and human--human agreement alongside judge scores to expose variation that a single pooled coefficient hides, with sensitivity checks across stratification cutoffs.
    \item \textbf{Use matched baselines and selection controls.} When interpreting pooled-label agreement, researchers should compare judges with a held-out human reference and apply selection controls suited to the panel size to rule out conditioning artifacts.
    \item \textbf{Pair HCA with stratified results.} HCA should be reported alongside stratified tables, bootstrap confidence intervals, and discriminative power $S$ to flag low-signal settings; scores above $1$ indicate closer proximity to consensus than the human reference rather than superior judgment of true quality.
    \item \textbf{Scope automation claims by consensus.} Claims about replacing human evaluators should be restricted to the tasks, criteria, and strata where reliability is established, while contested items are reported separately and retained for human review.
    \item \textbf{Release individual annotator ratings.} Benchmarks should publish raw per-rater scores together with aggregation rules and evaluation procedures so subsequent studies can analyze disagreement, build matched baselines, and test new protocols.
\end{enumerate}

\subsection{Threats to Validity}\label{sec:threats}

\noindent\textbf{Construct \& Internal Validity.} Our conclusion depends on how we operationalize ``human uncertainty.'' We mitigate this with three independent controls: a matched held-out-human reference, a split-panel design that defines and measures strata on disjoint raters, and a constant-quality-judge null. \textit{Choice of reference rater.} Because $\alpha_{\text{ref}}$ pools every held-out rater, no single rater drives it; replacing the pooled coefficient with the mean or median of the $K$ per-rater coefficients changes $\Delta_\alpha^{\text{match}}$ by at most $0.04$ on commit messages and $0.05$ on readability, with no sign change. On code summarization, with only three raters, the change is at most $0.11$ and flips the sign of one near-zero gap (Python Conciseness \textit{Partial}, $-0.07$ vs.\ $+0.04$). \textit{Choice of stratum thresholds.} The strata use benchmark-specific cutoffs (Section~\ref{subsec:human-uncertainty}). As a threshold-free check, we sort items by their mean pairwise human rating difference and compute $\Delta_\alpha^{\text{match}}$ in sliding windows of 100 (commit) or 120 (code summarization) items. The stratified pattern reappears: the matched gap is clearly positive in the high-consensus windows and shrinks toward zero in the most contested ones, turning negative there for some criteria. \textit{Held-out rater and strata.} In the matched gap, the held-out rater's own rating helps define the strata, and the judge faces the $K$-rater median rather than the $(K{-}1)$-rater median. As a check, we let the other $K{-}1$ raters define both the strata and the pooled label, and score the held-out rater and the judge against it. On commit messages, human agreement on \textit{High} items is then below $1$ ($0.93$ for \textit{What} and $0.76$ for \textit{Why}) with GPT-4 still significantly below it, and the \textit{Why}/\textit{Low} gap is $+0.04$ (95\% CI $[-0.03,+0.12]$). On readability, with strata formed from eight raters, the \textit{High}-stratum shortfall of Mistral Large falls from $+0.27$ to $+0.13$ and is no longer significant, and the lead of CodeVisionary over the held-out rater becomes significant over all items ($-0.16$). \textit{LLM stochasticity.} Where a source study averages repeated runs per item (three on commit messages; several on readability), our frontier LLM runs do the same. On code summarization, each judge is scored from a single sample per item. Our bootstrap CIs resample items, not judge samples, so they do not capture decoding variance.

\noindent\textbf{External \& Conclusion Validity.} Our findings may not generalize beyond the three SE benchmarks we study, or to annotator populations other than their panels of three to nine raters. To guard against over-claiming on small strata (e.g., the $3$--$12\%$ \textit{Low} strata in code summarization or the nine \textit{Low} snippets in readability) or low-signal benchmarks, we rely on FDR-controlled bootstrap tests, selection controls, and HCA's discriminative power $S$ with bootstrap CIs.

\section{Related Work}
\label{sec:relatedwork}

\noindent
\textbf{LLM-as-a-Judge for software engineering.}
As LLMs have become more capable, recent work has explored their use as automated evaluators.
He et al.~\cite{he2026llm} provide the first comprehensive review of this \emph{LLM-as-a-Judge} paradigm in software engineering.
Representative approaches include CodeJudge, which evaluates semantic correctness without test cases~\cite{tong2024codejudge}; Code-DiTing, which requires neither reference implementations nor executable tests~\cite{yang2025code}; and SE-Jury, which ensembles multiple judging strategies to improve alignment with human evaluation~\cite{zhou2025se}.
Other designs use structured criteria or intermediate reasoning, as in ICE-Score~\cite{zhuo2024ice}, HuCoSC~\cite{xu2024human}, and MCTS-Judge~\cite{wang2025mcts}, or adopt agentic and multi-evaluator frameworks, as in Agent-as-a-Judge~\cite{zhuge2024agent}, CodeVisionary~\cite{wang2025codevisionary}, and AIME~\cite{patel2024aime}.
These approaches build on broader multi-agent and panel-based evaluation~\cite{chan2024chateval,verga2024replacing}, while LLM judges have also been used for candidate reranking and iterative refinement~\cite{patel2024aime,zhou2025evaluating}.
Recent benchmarks further compare judge capabilities across code generation, repair, and test generation~\cite{jiang2026codejudgebench}.
Beyond generated code, LLM judges have been applied to artifacts involving human judgment, including commit messages~\cite{zeng2025evaluating}, code summaries~\cite{crupi2025effectiveness}, and bug-report summaries~\cite{kumar2024llms,wang2025can}.

\noindent
\textbf{Human baselines for evaluating LLM judges.}
Zeng et al.~\cite{zeng2025evaluating} evaluated GPT-4 on commit messages by correlating its scores with the mean ratings of six human evaluators.
They found that this correlation was close to the correlations between pairs of evaluators and interpreted the result as evidence of comparable performance.
This comparison leaves open how closely an individual human evaluator would agree with the pooled ratings of the other evaluators on the same messages.
Elangovan et al.~\cite{elangovan2025beyond} showed why the choice of reference matters. On disputed items, an automatic evaluator can agree more closely with the majority label than human raters agree with one another.
They therefore recommended reporting agreement separately by level of human consensus.
On our three SE benchmarks, we ask how closely a held-out human rater agrees with the rest of the panel within each consensus stratum.
This gives us a human baseline for interpreting the judge's agreement with the pooled label. We also test whether selecting items by human disagreement can itself produce an apparent judge advantage.

\section{Conclusion and Future Work}\label{sec:conclusion}

Across three SE benchmarks, the best judges reported by the original studies score below a matched human reference overall. On contested items, the choice of reference changes what an apparent judge advantage means. For commit-message \textit{Why}, GPT-4's lead shrinks when a held-out human is also scored against pooled ratings. For code-summary Content Adequacy, a lead persists after matching, but a human judge exhibits a gap of similar size when items are selected by rater disagreement. Neither result establishes that the model judges disputed items better than a human. Evaluating three frontier models and two multi-agent methods yields no consistent improvement over the released judges.

Under CSJE, we recommend reporting judge agreement alongside a matched human reference within each consensus stratum and checking how selection affects any apparent advantage on contested items. When a comparison also needs a single score, HCA summarizes proximity to individual human ratings while giving more weight to items that distinguish human agreement from random scoring. We recommend retaining human review for contested items. Future work should extend this evaluation to more SE tasks and to judges that explicitly model the distribution of human ratings.

\section{Data Availability}
Our replication package is available at \url{https://github.com/JundaHe/ICSE2027-When-Agreement-Hides-Disagreement}.

\bibliographystyle{IEEEtran}
\bibliography{reference}

\end{document}